\documentclass{aa}
 
\usepackage{graphicx}
\usepackage{txfonts} 
\def\ro{{\rm o}}
\def\re{{\rm e}}
\def\sss{{\sc superstructure}}
\def\Fe17{Fe~{\sc xvii}}
\newcommand{\eion}{({\rm e~+}~ion)\,}
\def\etal{{\it et\thinspace al.}\ }

\begin{document}
   \title{Atomic data from the Iron Project}
   \subtitle{LIII. Relativistic allowed and forbidden transition
probabilities for \Fe17}
\author{Sultana N. Nahar\inst{1}, Werner Eissner\inst{2}, Guo-Xin Chen\inst{1},
 and Anil K. Pradhan\inst{1}}
   \offprints{S.\,N. Nahar}
   \institute{Department of Astronomy, The Ohio State University,
              Columbus, OH 43210, USA\\
              \email{nahar@astronomy.ohio-state.edu}
\thanks{Complete electronic data tables of energies and 
trans\-ition probabilities are available from the CDS via anonymous ftp to 
{\tt cdsarc.u-strasbg.fr} (130.79.128.5) or
 via {\tt http://cdsweb.u-strasbg.fr/Abstract.html}
}
         \and
              Institut f\"ur Theoretische Physik, Teilinstitut 1,
              70550 Stuttgart, Germany}
\abstract{An extensive set of fine structure levels and corresponding 
transition probabilities for allowed and forbidden transitions in 
\Fe17\ is presented. A total of 490 bound energy levels of \Fe17\ of 
total angular momenta 0 $\leq J \leq $ 7 of even and 
odd parities with 2 $\leq n\leq $ 10, 0 $\leq l\leq $ 8, 0 $\leq L\leq$ 8,
and singlet and triplet multiplicities, are obtained. They translate to
over 2.6$\times 10^4$ allowed (E1) transitions that are of dipole and 
intercombination type, and about 3000 forbidden transitions that
include electric quadrupole (E2), magnetic dipole (M1), electric 
octopole (E3), and magnetic quadrupole (M2) type representing the most 
detailed calculations to date for the ion. Oscillator strengths $f$, line 
strengths $S$, and coefficients $A$ of spontaneous emission for the E1 
type transitions are obtained in the relativistic Breit-Pauli R-matrix 
approximation. $A$ valus for the forbidden transitions are obtained from 
atomic structure calculations using codes SUPERSTRUCTURE and GRASP. The 
energy levels are identified in spectroscopic notation with the help of a 
newly developed level identification algorithm. {\em Nearly\/} all 52 
spectroscopically observed levels have been identified, their binding 
energies agreeing within 1\% with our calculation.  Computed transition 
probabilities are compared with other calculations and measurement. The 
effect of 2-body magnetic terms and other interactions is discussed. 
Present data set enhances by more than an order of magnitude the heretofore 
available data for the transition probabilities of \Fe17.
}
\authorrunning{S.\,N. Nahar \etal}
\titlerunning{Transition probabilities for \Fe17}
\maketitle
\section{Introduction}

Ne-like \Fe17\ attracts great astrophysical interest with some of the 
most prominent spectral lines in the X-ray and the EUV regimes. These 
lines are abundantly evident from diverse sources such as the solar 
corona and other stellar coronae (e.\,g.\ Brickhouse \etal 2001), and 
active galactic nuclei (e.\,g.\ Lee \etal 2001). \Fe17\ also plays a role 
in benchmarking laboratory experiments and theoretical calculations. 
Recent Iron Project (IP, Hummer et al. 1993) work has included the 
computation of collision strengths and rate coefficients by electron 
impact excitation of \Fe17\, and diagnostics of laboratory and
astrophysical spectra (Chen and Pradhan 2002; Chen, Pradhan and Eissner 
2002 -- hereafter CPE02). Spectral analysis moreover requires transition
probabilities for observed allowed and forbidden transitions. Transition 
probablities are also required to account for radiative cascades from 
higher levels that contribute to level populations; cascades
generally proceed via strong dipole allowed transitions, and may
entail fairly highly excited levels. Therefore a fairly large and
complete set of data is needed for astrophysical models of \Fe17.

Smaller sets of transitions are available from other sources. An 
evaluated compilation of data, obtained by various investigators using 
different approximations, can be found in the National Institute for 
Standards and Technology database (NIST: www.nist.gov). A previous set 
of non-relativistic data for \Fe17\ was obtained by M.\,P. Scott 
under the Opacity Project (OP 1995, 1996), which are accessible through 
the OP database, TOPbase (Cunto \etal 1993). These results are in LS 
coupling and consider only the dipole allowed $LS$ multiplets; no 
relativistic effects are taken into account.

The present calculations are carried out for extensive sets of oscillator
strengths, line strengths, and transition probabilities of dipole allowed,
intercombination, and forbidden electric quadrupole and octopole,
magnetic dipole and quadrupole fine structure (FS) transitions in
\Fe17\ up to $n \leq 10$. Transitions of type E1 are obtained in the 
relativistic Breit-Pauli R-matrix method developed under the Iron Project.
Configuration mixing type atomic structure calculations, using codes
SUPERSTRUCTURE (Eissner \etal 1974) and GRASP (Parpia \etal 1996) which is
based upon the multiconfiguration Dirac-Fock (MCDF) method, are employed 
for the forbidden E2, E3, and M1, M2 transitions. One of the primary tasks 
is the spectroscopic identification of levels and lines of E1 transitions.
We apply the recently developed techniques (Nahar and Pradhan 2000)
for a reasonably complete spectroscopic dataset to \Fe17.

\section{Formulation}

We employ the relativistic Breit-Pauli {\it R}-matrix (BPRM) approach in 
a collision type calculation for bound states followed by computing 
radiative processes: Scott and Burke 1980, Scott and Taylor 1982, Hummer 
\etal 1993, Berrington \etal 1995. Unlike calculations in $LS$ coupling, 
when radiative transition amplitudes vanish unless $\Delta S=0$, 
intermediate coupling calculations include intercombination lines.

Details of this close coupling (CC) approach to radiative processes are
discussed in earlier papers, such as in the first large scale relativistic 
BPRM calculations for bound-bound transitions in Fe{\sc~xxiv} and 
Fe{\sc~xxv} (Nahar and Pradhan 1999), Fe{\sc~v} (Nahar \etal 2000), 
Ar XIII and Fe XXI (Nahar 2000). However, in the present work, 
electric octopole and magnetic dipole transitions are included for the 
first time in the IP series. A brief outline for the formulations is 
henceforth given.

The wavefunction $\mit\Psi(E)$ for the ($N$\,+\,1) electron system with total
spin and orbital angular momenta symmetry $SL\pi$ or total angular momentum
symmetry $J\pi$ is expanded in terms of `frozen' $N$-electron target ion
functions $\chi_i$ and vector coupled collision electrons $\theta_i$,
\begin{equation}
\mit\Psi_E\eion = {\cal A}\sum_i \chi_i({\it ion})\theta_i
 + \sum_j c_j \Phi_j\eion,
\end{equation}
in some specific state $S_iL_i\pi_i$ or level $J_i\pi_i$, index $i$ marking
channels $S_iL_i(J_i)\pi_i \ k_i^2\ell_i(SL\pi$~or~$ J\pi)$ with energy
$k_i^2$ of the colliding electron. The second sum expands correlation
functions $\mit\Phi_j$ as products with $N$\,+\,1 bound orbital functions
that (a) compensate for the orthogonality conditions between the continuum and
the bound orbitals, and (b) represent additional short-range correlation
that is often of crucial importance in scattering and radiative CC
calculations for each $SL\pi$.

In IP work we restrict the ($N$\,+\,1)-electron Breit-Pauli Hamiltonian to
\begin{equation}
H_{N+1}^{\rm BP}=H_{N+1}^{\rm NR}+H_{N+1}^{\rm mass} + H_{N+1}^{\rm Dar}
+ H_{N+1}^{\rm so},
\end{equation}
where $H_{N+1}^{\rm NR}$ is the non-relativistic Hamiltonian
\begin{equation}
H_{N+1}^{\rm NR} = \sum_{i=1}\sp{N+1}\left\{-\nabla_i\sp 2 - \frac{2Z}{r_i}
        + \sum_{j>i}\sp{N+1} \frac{2}{r_{ij}}\right\} ;\label{HNR}
\end{equation}
among the three trailing relativistic terms mass-velocity correction,
$H^{\rm mass} = -{\alpha^2\over 4}\sum_i{p_i^4}$, and Darwin term,
$H^{\rm Dar} = {Z\alpha^2 \over 4}\sum_i{\nabla^2({1\over r_i})}$. do 
not break $LS$ symmetry while improving energy positions, whereas the 
spin-orbit interaction, $H^{\rm so}= Z\alpha^2 \sum_i{1\over r_i^3}
{\bf l_i.s_i}.$, splits terms $LS$ into fine-structure levels labelled 
by $J\pi$. This Hamiltonian does not include two-body spin-spin, mutual 
spin-orbit and spin-other-orbit terms nor non-fine structure 
contributions such as orbit-orbit interaction.

{\it R}-matrix solutions of coupled equations to total symmetries $LS$
are recoupled in a pair coupling scheme on adding spin-orbit interaction 
to obtain (e + {\it ion}) states of total $J\pi$, in the end yielding 
($N$\,+\,1)-electron solutions
\begin{equation}
H^{\rm BP}_{N+1}\mit\Psi = E\mit\Psi\,.
\end{equation}
Rather than dealing with positive energies ($E\,>\,0$) as in ordinary 
collision processes we focus on an eigenvalue problem ($E\,<\,0$) for 
the electron described by $\theta$, leading to discrete bound states 
$\mit\Psi_{\rm B}$.

The transition matrix element for radiative bound-bound excitation
or de-excitation can be reduced to the line strength as
\begin{eqnarray}
S^{{\rm X}\lambda}(ij)&=&
 \Big|\big\langle{\mit\Psi}_j\big\Vert O^{{\rm X}\lambda}\big\Vert
 {\mit\Psi}_i\big\rangle\Big|^2\,,\qquad\quad S(ji)=S(ij).\label{eq:Xlam}
\end{eqnarray}
For electric (X=E) multipole transitions in the length formulation (and 
long wave-length approximation), 
\begin{eqnarray}
 O^{{\rm E}\lambda} = b^{[\lambda]}\sum_{p=1}^{N+1}
 {\rm C}^{[\lambda]}(p)r_p^{\lambda}, \ \
 b^{[\lambda]} = \sqrt{\frac{2}{\lambda+1}}
\label{eq:Elam}
\end{eqnarray}
Transition probabilities $A$ and absorption oscillator strengths 
($f$-values) between bound states $i$ and $j$ and excitation energy 
$E_{ij} = E_j-E_i$ are written in terms of the line strength $S$, 
observing that Eq.\,(\ref{HNR}) implies scaling of energies in units of
Ry\,$\displaystyle{= \frac{\alpha^2}{2}m_{\rm el}c^2 = 13.6}$\,eV,
hence time unit $\tau_0 = \hbar/{\rm Ry} = 4.838\cdot 10^{-17}\cdot$\,s:
\begin{eqnarray}
f_{ij}&=&{E_{ji}\over {3g_i}}S^{\rm E1}(ij),
\quad \ g_if_{ij}=-g_jf_{ji} \ = \ (gf)_{ij}
\label{eq:fij}\\
A_{ji}^{\rm E1}\cdot\tau_0&=&\alpha^3{g_i\over g_j}E_{ji}^2f_{ij}\label{eq:Aji}
\end{eqnarray}
in the case of electric dipole radiation E$\lambda =\,$E1.
The symbols in these equations have their usual meaning, in particular 
$g_j$ and $g_i$ being the statistical weights of the upper and lower 
states respectively. Hypervirial identities arising from the commutator 
$[rH]_-$ yield alternative formulations, velocity formulation for a start, 
that probe the radial wave functions less far out. With $H^{\rm NR}$ it 
leads to simple substitutions of $r^\lambda$ in eq.\,(\ref{eq:Elam}) --- 
but to additional terms of order $\alpha^2$ for $H^{\rm BP}$! BPRM 
ignores such `velocity' terms: they are not large enough though for 
\Fe17\ to render comparison of length with velocity results a useless 
tool (yet better left to NR-results). In the magnetic dipole case the 
radiative operator to the line strength expression
(\ref{eq:Xlam}) reads
\begin{equation}
O^{\rm M1}=\sum_p
 {\rm l}(p)+2{\rm s}(p)+\alpha^2\bigg\{\frac{\partial^2\ }{\partial r_p^2}
 +\dots +\sum_{p'>p}\frac{\dots}{r_{p'p}}\bigg\}\,;
\label{eq:OM1}
\end{equation}
where the sum runs over electron coordinates, $l$ and $s$ are the orbital
and spin operators respectively. Details on the correction of relative BP order can be found in {\sc O ii}
work of 1981 by Eissner and Zeippen.
Magnetic quadrupole ($\lambda$=2) radiation is treated to lowest order, i.\,e.
\begin{equation}
O^{{\rm M}\lambda} = b^{[\lambda]}\sum_pr_p^{\lambda-1}\Big[
 {\rm C}^{[\lambda-1]}(p)\times\big\{
 {\rm l}(p)+(\lambda+1){\rm s}(p)\big\}\Big]^{[\lambda]}\!. \label{eq:Mlam}
\end{equation}

The lifetime of a level can be computed as
\begin{eqnarray}
\tau&=&\frac{1}{A_k}\,,\\
{\rm where}\qquad
 A_k&=&\sum_i A_{ki}\,\nonumber
\end{eqnarray}
is the total radiative transition probability for
level $k$, i.\,e.\
\begin{equation}
g_iA_{ki}^{\rm E1}=2.6774\times 10^9{\rm s}^{-1}\,(E_i-E_k)^3S^{\rm E1}(i,k)
\end{equation}
(the observed rate) in the electric dipole case E1. The Einstein coefficients
for spontaneous decay by higher order multipole radiation that need be
considered for transitions down to the 10\,\AA\ range read as follows:\\[.5ex]
electric quadrupole (E2) and magnetic dipole (M1)
\begin{eqnarray}
g_jA^{\rm E2}_{ji}&=&2.6733\times 10^3{\rm s}^{-1}\,(E_j-E_i)^5S^{\rm E2}(i,j)
 \label{eq:E2}\\
\mbox{and}\nonumber\\
g_jA^{\rm M1}_{ji}&=&3.5644\times 10^4{\rm s}^{-1}\,(E_j-E_i)^3S^{\rm M1}(i,j);
 \label{eq:M1}
\end{eqnarray}
electric octopole (E3) and magnetic quadrupole (M2)
\begin{eqnarray}
g_jA^{\rm E3}_{ji}&=&1.2050\times 10^{-3}{\rm s}^{-1}\,(E_j-E_i)^7S^{\rm E3}
(i,j)\label{eq:E3}\\\mbox{and}\nonumber\\
g_jA^{\rm M2}_{ji}&=&2.37268\times 10^{-2}{\rm s}^{-1}\,(E_j-E_i)^5S^{\rm M2}
(i,j)\,.\label{eq:M2}
\end{eqnarray}
In approximations like BP one should be careful with the radiative magnetic
operators about terms of order $\alpha^2$, in particular in $O^{\rm M1}$, which
cannot connect different configurations by its leading term $l+2s$ because 
the (tensor-) radial portion reduces to trivial 1; \sss\ does add both 1-body
and 2-body contributions of Breit-Pauli order to M1 but not to M2.

\section{Computation}

BPRM calculations span several stages of computation (Berrington \etal 
1995). We take radial Fe{\sc~xviii} wavefuntions from \sss\ (Eissner 
\etal 1974) as input to {\tt STG1} to compute Slater, magnetic and 
multipole integrals --- obtained with Thomas-Fermi scaling parameters 
$\lambda_{nl}$ of 1.3835, 1.1506, 1.0837, 1.0564, 1.0175, 1.0390 for 
orbitals $nl=$ 1s, 2s, 2p\ldots\,3d, which leads to excited levels 
2s$^2$2p$^5\,^2$P$^\ro_{1/2}$ and 2s2p$^6\,^2$S$_{1/2}$ at 0.9403 and 
9.8092 Rydbergs above the ground state 2s$^2$2p$^5\,^2$P$^\ro_{3/2}$ 
(while including correlation terms from 6 configurations: 2s$^2$2p$^43l$ 
and 2s2p$^53l$ --- `1s$^2$' suppressed for brevity); the excitation 
energies above the ground state compare with NIST data of 0.93477 and 
9.7023\,Ry respectively. Other excited levels of Fe{\sc~viii} lie too 
high to play a role as parent for any \Fe17\ bound states (50\,Ry
separating M- from L-shell: level 2s$^2$2p$^4$3s\,$^4$P$_{5/2}$ at 
57.01\,Ry), and therefore need not be considered for radiative
calculations. Radial integrals for the partial wave expansion in Eq.\,1 
are specified for orbitals $0\leq\ell\leq 9$ as a basis of {\tt NRANG2 = 
11} `continuum' functions --- sufficient for bound electrons with $n < 10$ 
at a radius {\tt RA = 2.3750} [Bohr radii $a_0$] of the $R$-matrix box.

Along with the target description {\tt STG2} input specifies which
collisional \Fe17\ symmetries $LS$ eventually contribute to 0 
$\leq J\leq$ 7 or 8 of even and odd parities, namely $0\leq L \leq$ 7 
or 8, and multiplicities $(2S+1)$=1, 3. The second term in Eq. (1), on
bound state correlation functions, is specified to include all possible 
($N$\,+\,1)-particle configurations from a vacant 2s shell to maximum 
occupancies 2s$^2$, 2p$^6$, 3s$^2$, 3p$^2$, and 3d$^2$.

Stage {\tt RECUPD} transforms to collisional symmetries $J\leq 7$ or 8
in a pair-coupling representation, and the \eion Hamiltonian $R$-matrices
for each total $J\pi$ are diagonalized in {\tt STGH} employing
{\rm observed\/} target energies.

In {\tt STGB} fine structure bound levels are found through the poles 
in the (e + ion) Hamiltonian, searched over a fine mesh of effective 
quantum number $\nu$: $\Delta\nu$ = 0.001. The mesh is orders 
of magnitude finer than the typical $\Delta\nu$ = 0.01 required to find 
$LS$ energy terms. Intermediate coupling calculations therefore need 
orders of magnitude more CPU time than calculations in $LS$ coupling. 
Since the fine structure components of higher excited states are more 
densely packed, a mesh finer than $\Delta\nu$ = 0.001 is essential to 
avoid missing any levels. 

Spectroscopically identifying a large number of fine structure levels poses
a major challenge, as the BP Hamiltonian is labelled only by the total
angular momentum and parity, i.\,e.\ by $J\pi$, which is incomplete for
unique identification. Complete identification of levels is needed 
for various spectral diagnostics and spectrocopic applications in a lab. 
A new procedure has been 
developed and encoded in the program PRCBPID to identify these levels 
by a complete set of quantum numbers through analysis of coupled channels 
in the CC expansion (Nahar \& Pradhan 2000). This procedure generally 
yields unambiguous level identification for most levels. However, for 
mixed levels where the identification is to some extent arbitrary, we 
assign levels in descending multiplicity $(2S+1)$ and total angular 
orbital momentum $L$. The full
spectroscopic designation reads $C_t(S_tL_t\pi_t)J_tnlJ(SL)\pi$,
where $C_t$, $S_tL_t\pi_t$, $J_t$ are the configuration, parent term and
parity, and total angular momentum of target states, $nl$ are the principal
and orbital quantum numbers of the outer or valence electron, and $J$
and $SL\pi$ are the total angular momentum, term and parity of the
($N$+1)-electron system. The procedure also establishes a correspondence
between the fine structure levels and their proper $LS$ terms, and enables
completeness checks to be performed as exemplified below.

{\tt STGBB} can compute radiative data for transitions of type E1 and E2; 
the code exploits methods developed by Seaton (1986) to evaluate the outer 
region ($>{\tt\,RA}$) contributions to the radiative transition matrix 
elements. However, present work reports only E1 transitions from STGBB.
Results for other types of transitions are obtained from \sss, first 
optimizing the energy functional over the lowest 49 terms $LS$ (Chen 
\etal 2002, CPE02). They arise from 15 configurations: 2s$^2$2p$^6$, 
2s$^2$2p$^53l$, 2s$^2$2p$^54l$, 2s$^1$2p$^63l$, and 2s$^1$2p$^64l$; the 
scaling parameters $\lambda _{nl}$ for the Thomas-Fermi-Dirac-Amaldi 
type potential of orbital $nl$ are listed in table~1 of CPE02. Much effort 
was devoted to choosing scaling parameters to optimise the target 
wavefunctions of the M-shell levels. The primary criteria in this 
selection are agreement with the observed values for (a) level energies 
and fine structure splittings within the lowest terms $LS$, and (b) 
$f$-values for a number of the low lying dipole allowed transitions. 
Another practical criterion is that the calculated coefficients
$A$ should be variationally stable.

Experimental energy level differences are employed in the calculation
of all types of transition probabilities wherever available, ensuring proper
phase space (or energy) factors for $f$ or $A$; only a
small number of \Fe17\ levels are spectroscopically observed though.

In addition to  over 26,000 electric dipole trans\-itions we have 
computed $A^{\rm E2}$, $A^{\rm M1}$, $A^{\rm E3}$ and $A^{\rm M2}$ for 
more than 3000 forbidden transitions among the first 89 levels. About half 
of the computed forbidden transition probabilities are larger than 
$10^3$s$^{-1}$.  Selected transtions (Table~7) are compared with various 
other calculations. Results by Safronova \etal (2002, private 
communication) are included for comparison.

\section{Results}

We first describe the BPRM calculations for the energy levels and E1 dipole 
and intercombination transitions in \Fe17\ and then discuss higher multipole
order radiation.

\subsection{Fine structure levels}
A total of 490 bound fine structure energy levels of \Fe17\ are obtained
from interacting channels, or Rydberg series
\begin{equation}
E = E_t - \frac{z^2}{\nu^2}\,,\qquad\quad \nu=n-\mu_{l\pm 1/2}(t)\label{eq:RyS}
\end{equation}
with series limits $E_t$ at the 3 Fe{\sc~xviii} `target' levels 
2s$^2$2p$^5$\,$^2$P$^\ro_{3/2,\ 1/2}$, 2s2p$^6\,^2$S$_{1/2}$, for symmetries
0 $\leq J \leq$ 7 (both parities), implying series orbitals 0 $\leq l\leq $ 8.
In intermediate coupling language we consider bound state levels of \Fe17\
to angular momenta $L \leq$ 8 of singlet and triplet symmetries (multiplets
to high $L$ may thus be incomplete). Series are kept below effective quantum
numbers $\nu = 11$ measured from the target ground state. 
These are the most
detailed close coupling calculations to date for the ion.

Table 1 tentatively matches the 52 spectroscopically observed levels from 
NIST with identified levels from our calculations (the level index $I\!_J$, 
in ascending energy order within a given symmetry $J\pi$, is most useful 
for reference in subsequent tables). Calculated effective quantum numbers 
$\nu_{\rm c}$ of the first 14 entries differ from observation within 
numerical uncertainties and errors due to neglect of two-body magnetic 
effects: typically $\Delta\mu  \equiv  \Delta\nu=.0005$. The abrupt jump 
to .0027 at level 15 and typical values of 0.002 thereafter can be 
explained by the effect of M-shell target levels, for good reasons not 
included in our calculation. Results from SUPERSTRUCTURE for 
Fe{\sc~xviii} reveal that the set of M-shell levels, comprising 105 levels, 
begins at 57.08\,Ry above the Fe{\sc~xviii} ground state, and with the
binding energy of 92.76\,Ry for a 2p electron, from the first entry of
Table 1, a first quasi-degenerate state lies an adequate 35.68\,Ry below 
the ground state.  We see that such homologous states do not seriously 
affect the accuracy of our calculation. More important is that M-shell 
target configurations do not render it incomplete: a binding energy of 
about 40\,Ry for a 3s electron taken from entries 2--5 of Table 1 would 
lead to true new levels beginning (60-40)\,Ry {\em above\/} the 
ionization limit. It is also worth noting at 2--5
entries that the 4 quantum defects are close enough for mere differences 
in the Coulomb environment, as s-electrons are not affected by ordinary 
spin-orbit coupling. Way down the table agreement deteriorates. While 
$\Delta\mu\approx 0.005$ may be considered acceptable and a value 0.01 
needs some explanation, the attempts with the 7d and more so 8d levels 
are an utter failure, 8d off by 0.13 and 0.04, not to speak of a negative 
`observed' quantum defect of the second 8d level. Such binding energies 
$E_{\rm o}$ are unlikely.
\begin{table}
\noindent{Table 1. Comparing effective quantum numbers $\nu_{\rm o}$ of
observed binding energies $E_{\rm o}$ with $\nu_{\rm c}$ computed in stage
{\tt STGB} of BPRM ($\nu$ measured from respective Fe{\sc~xviii} threshold
$t$). Index $I\!_J$ counts levels within symmetry $J\pi$ in energy order,
* indicating that level $J$ belongs to an incompletely observed multiplet.
\\[.8ex]}
\begin{tabular}{l@{\hspace{.6em}}l@{\hspace{.7em}}l@{\hspace{.7em}}rl
@{\hspace{.9em}}c@{\hspace{.7em}}c@{\hspace{.6em}}c@{}}
\hline
\noalign{\smallskip}
\multicolumn{2}{c}{\it level} &$J$ &$I\!_J$ &\multicolumn{1}{c}{$E_{\rm o}$/Ry}
 & $\nu_{\rm o}$ & $\nu_{\rm c}$ &$t$\\
            \noalign{\smallskip}
\hline
            \noalign{\smallskip}
2s22p6           &$^1$S  &     0 &  1 &92.760 &1.7651 &1.7643 &1\\
2s22p5(2P*3/2)3s &$^3$P$^\ro$& 2 &  1 &39.463 &2.7062 &2.7063 &1\\
2s22p5(2P*3/2)3s &$^3$P$^\ro$& 1 &  1 &39.323 &2.7110 &2.7111 &1\\
2s22p6(2P*1/2)3s &$^3$P$^\ro$& 0 &  1 &38.533 &2.7060 &2.7064 &2\\ 
2s22p6(2P*1/2)3s &$^1$P$^\ro$& 1 &  2 &38.446 &2.7090 &2.7095 &2\\
2s22p53p         &$^3$S    &   1 &  1 &37.238 &2.7858 &2.7858 &1\\
2s22p53p         &$^3$D    &   3 &  1 &36.863 &2.8000 &2.8001 &1\\
2s22p53p         &$^3$D    &   2 &  1 &36.981 &2.7955 &2.7958 &1\\
2s22p53p         &$^3$D    &   1 &  3 &36.093 &2.7937 &2.7945 &2\\
2s22p53p         &$^1$P    &   1 &  2 &36.780 &2.8031 &2.8034 &1\\ 
2s22p53p         &$^3$P    &   2 &  2 &36.646 &2.8082 &2.8085 &1\\
2s22p53p         &$^3$P    &   1 &  4 &35.854 &2.8028 &2.8034 &2\\
2s22p53p         &$^3$P    &   0 &  2 &36.244 &2.8238 &2.8246 &1\\
2s22p53p         &$^1$D    &   2 &  3 &35.826 &2.8039 &2.8046 &2\\[.4ex] 
2s22p53p         &$^1$S    &   0 &  3 &34.871 &2.8410 &2.8437 &2\\
2s22p53d         &$^3$P$^\ro$& 2    & 2 &33.662 &2.9301 &2.9324 &1\\
2s22p53d         &$^3$P$^\ro$& 1    & 3 &33.778 &2.9250 &2.9260 &1\\ 
2s22p53d         &$^3$P$^\ro$& 0    & 2 &33.862 &2.9214 &2.9226 &1\\ 
2s22p53d         &$^3$F$^\ro$& 4    & 1 &33.656 &2.9303 &2.9329 &1\\ 
2s22p53d         &$^3$F$^\ro$& 3    & 1 &33.599 &2.9329 &2.9346 &1\\ 
2s22p53d         &$^3$F$^\ro$& 2    & 4 &32.672 &2.9325 &2.9346 &2\\ 
2s22p53d         &$^1$D$^\ro$& 2    & 3 &33.472 &2.9384 &2.9403 &1\\ 
2s22p53d         &$^3$D$^\ro$& 3    & 2 &33.393 &2.9419 &2.9444 &1\\ 
2s22p53d         &$^3$D$^\ro$& 2    & 5 &32.598 &2.9357 &2.9380 &2\\ 
2s22p53d         &$^3$D$^\ro$& 1    & 4 &33.052 &2.9570 &2.9595 &1\\ 
2s22p53d         &$^1$F$^\ro$& 3    & 3 &32.563 &2.9373 &2.9397 &2\\ 
2s22p53d         &$^1$P$^\ro$& 1    & 5 &32.070 &2.9591 &2.9525 &2\\ 
2s2p63p          &$^3$P$^\ro$& 1\,* & 6 &27.159 &2.8000 &2.8047 &3\\ 
2s2p63p          &$^1$P$^\ro$& 1    & 7 &26.836 &2.8124 &2.8171 &3\\
2s22p5(2P*3/2)4s &$^3$P$^\ro$& 1\,* & 8 & 20.899 &3.7187 &3.7209 &1\\
2s22p5(2P*1/2)4s &$^1$P$^\ro$& 1    & 9 & 20.014 &3.7142 &3.7188 &2\\
2s22p54d         &$^3$P$^\ro$& 1\,* &10 & 18.802 &3.9205 &3.9283 &1\\
2s22p54d         &$^3$D$^\ro$& 1\,* &11 & 18.455 &3.9572 &3.9623 &1\\
2s22p54d         &$^1$P$^\ro$& 1    &12 & 17.590 &3.9498 &3.9540 &2\\
2s22p5(2P*3/2)5s &$^3$P$^\ro$& 1\,* &13 & 12.960 &4.7222 &4.7201 &1\\
2s22p5(2P*1/2)5s &$^1$P$^\ro$& 1    &14 & 12.022 &4.7228 &4.7173 &2\\
2s22p55d         &$^3$P$^\ro$& 1\,* &15 & 12.022 &4.9030 &4.9258 &1\\
2s22p55d         &$^3$D$^\ro$& 1\,* &16 & 11.776 &4.9539 &4.9610 &1\\

2s22p55d         &$^1$P$^\ro$& 1    &17 & 10.910 &4.9395 &4.9484 &2\\
2s2p64p          &$^3$P$^\ro$& 1\,* &18 & 10.236 &3.8072 &3.8142 &3\\
2s2p64p          &$^1$P$^\ro$& 1    &19 & 10.090 &3.8212 &3.8235 &3\\
2s22p5(2P*3/2)6s &$^3$P$^\ro$& 1\,* &20 &\ 8.7776 &5.7380 &5.7196 &1\\ 
2s22p5(2P*3/2)6d &$^3$P$^\ro$& 1\,* &22 &\ 8.1488 &5.9555 &5.9547 &1\\ 
2s22p5(2P*1/2)6d &$^1$P$^\ro$& 1\,* &24 &\ 7.2558 &5.9401 &5.9417 &2\\ 
2s22p5(2P*3/2)7s &$^3$P$^\ro$& 1\,* &25 &\ 6.3810 &6.7298 &6.7220 &1\\
2s22p5(2P*3/2)7d &$^3$P$^\ro$& 1\,* &26 &\ 5.9709 &6.9571 &6.9240 &1\\
2s22p5(2P*1/2)7d &$^1$P$^\ro$& 1\,* &29 &\ 5.0232 &6.9647 &6.9422 &2\\
2s22p5(2P*3/2)8d &$^3$P$^\ro$& 1\,* &31 &\ 4.4582 &8.0514 &7.9267 &1\\
2s22p5(2P*1/2)8d &$^1$P$^\ro$& 1\,* &35 &\ 3.6016 &7.9817 &7.9397 &2\\
2s2p65p          &$^3$P$^\ro$& 1\,* &42 &\ 2.7450 &4.8185 &4.8141 &3\\
2s2p65p          &$^1$P$^\ro$& 1    &43 &\ 2.7450 &4.8185 & 4.8250 &3\\[.6ex]
\hline      \noalign{\smallskip}
\multicolumn{8}{r}{n.\,b.\ $E_t$/Ry = 0.0, 0.9348, 9.7023\
[M-shell: 57.08,\,\ldots\,74.14,}\\  
\multicolumn{8}{r}{N-shell: 77.05,\,\ldots\,91.36,
                   O-shell: 85.71,\,\ldots\,98.66]}\\[.6ex]
\hline
\end{tabular}
\end{table}

A complete set of energy levels to \Fe17\ is available electronically.
As in recent work (e.\,g.\ Nahar \etal 2000) the energies are
presented in two formats: (i) in $LS$ term order for spectroscopy and
completeness check, and (ii) in $J\pi$ order for practical applications.
In the term format (i) the fine structure components of a $LS$ term are
grouped together according to the same configuration, useful for
spectroscopic diagnostics. It also checks for completeness
of a set of energy levels that should belong to same $LS$ value
and detects any missing level. Table 2a presents a sample of the table
containing total sets of energies. The table contains partial sets of
levels of \Fe17. The columns specify the core $C_ t(SL\pi~J)_t$, the label
$nl$ of the outer electron, total angular momentun $J$, energy in Rydbergs,
the effective quantum number $\nu$ of the valence electron, and possible
term designations $LS$ of the level.
No effective quantum number is assigned to an equivalent electron state.

The top line of each set in Table 2a gives the number {\tt Nlv} of expected
fine structure levels, spin and parity of the set ($^{2S+1}L^\pi$), and
the values of $L$; the total angular quantum numbers $J$ associated with 
each $L$ are quoted parenthetically. This line is followed by the set of 
BPRM energy levels of same configurations. {\tt Nlv(c)}, at the end of 
the set, specifies the total number of $J$-levels obtained. If Nlv = 
Nlv(c) for a set, the calculated energy set is complete. Correspondence 
of couplings and completeness of levels is established by the program 
PRCBPID, which detects and prints missing levels. Each level of a set is 
further identified by all possible terms $LS$ (specified in the last column 
of the set). Multiple $LS$ terms are arranged according to multiplicity 
$(2S+1)$ and $L$ as mentioned above. It may be noted that levels are 
grouped consistently, closely spaced in energies and effective quantum 
numbers, confirming proper designation of terms $LS$. The effective
quantum number ($\nu$) is expressed upto two significant digits after 
the decimal point; the main object is to show the consistency of fine
structure components in the $LS$ grouping. Each level may be assigned to 
one or more $LS$ terms in the last column. For a multiple designation 
Hund's rule of decreasing multiplicity $(2S+1)$ and $L$ is applied for 
further arrangement. One reason for specifying all possible terms is that 
the order of calculated and measured energy levels may not exactly match. 
Another reason is that
although our term order arrangement may not apply to all cases
for complex ions, it is nonetheless useful in order to establish
completeness of fine structure components of a given $LS$ multiplet.
\begin{table}
\noindent{Table 2a. Sample table of fine structure energy levels of \Fe17\
as sets of $LS$ term components; $C_t$ is the core configuration, 
$\nu$ is the effective quantum number.\\
}
\begin{tabular}{l@{\hspace{.5em}}c@{\hspace{.5em}}rrrcrl}
\noalign{\smallskip}
\hline
\noalign{\smallskip}
 \multicolumn{2}{c}{$C_t(S_tL_t\pi_t)$} & $J_t$
 & $nl$ & $J$ & $E$/Ry & $\nu$ & $SL\pi$\\
\noalign{\smallskip}
 \hline
\noalign{\smallskip}
 \multicolumn{8}{l}{ Eqv electron/unidentified levels, parity: e
          }\\
 &     &    &   &  0& $-$92.8398 &  0.00 &  1~ S e \\
 \noalign{\smallskip}
 \multicolumn{8}{l}{Nlv(c)=~ 1~ : set complete   } \\
 \noalign{\smallskip}
 \hline
 \noalign{\smallskip}
 \multicolumn{8}{l}{Nlv=~ 3,~~$^3\!L^\ro$:
 P ( 2 1 0 )                                                             } \\
 \noalign{\smallskip}
 2s22p5   &(2Po)& 3/2& 3s&  2& $-$39.4577 &  2.71 &  3~ P o       \\
 2s22p5   &(2Po)& 1/2& 3s&  1& $-$39.3187 &  2.68 &  3~ P o       \\
 2s22p5   &(2Po)& 1/2& 3s&  0& $-$38.5208 &  2.71 &  3~ P o       \\
 \noalign{\smallskip}
 \multicolumn{8}{l}{Nlv(c)=~ 3~ : set complete
                                                                       } \\
 \noalign{\smallskip}
 \hline
 \noalign{\smallskip}
 \multicolumn{8}{l}{Nlv=~ 1,~~$^1\!L^\ro$:
 P ( 1 )                                                                 } \\
 \noalign{\smallskip}
 2s22p5   &(2Po)& 3/2& 3s&  1& $-$38.4324 &  2.74 &  1~ P o       \\
 \noalign{\smallskip}
 \multicolumn{8}{l}{Nlv(c)=~ 1~ : set complete
                                                                       } \\
 \noalign{\smallskip}
 \hline
 \noalign{\smallskip}
 \multicolumn{8}{l}{Nlv=~ 7,~~$^3\!L^{\re}$:
 S ( 1 ) P ( 2 1 0 ) D ( 3 2 1 )                                         } \\
 \noalign{\smallskip}
 2s22p5   &(2Po)& 1/2& 3p&  1& $-$37.2397 &  2.75 &  3~ SPD e     \\
 2s22p5   &(2Po)& 1/2& 3p&  2& $-$36.9744 &  2.76 &  3~ PD e      \\
 2s22p5   &(2Po)& 3/2& 3p&  3& $-$36.8541 &  2.80 &  3~ D e       \\
 2s22p5   &(2Po)& 3/2& 3p&  2& $-$36.6391 &  2.81 &  3~ PD e      \\
 2s22p5   &(2Po)& 3/2& 3p&  0& $-$36.2221 &  2.82 &  3~ P e       \\
 2s22p5   &(2Po)& 1/2& 3p&  1& $-$36.0724 &  2.79 &  3~ SPD e     \\
 2s22p5   &(2Po)& 3/2& 3p&  1& $-$35.8374 &  2.84 &  3~ SPD e     \\
 \noalign{\smallskip}
 \multicolumn{8}{l}{Nlv(c)=~ 7~ : set complete
                                                                       } \\
 \noalign{\smallskip}
 \hline
 \noalign{\smallskip}
 \multicolumn{8}{l}{Nlv=~ 3,~~$^1\!L^{\re}$:
 S ( 0 ) P ( 1 ) D ( 2 )                                                 } \\
 \noalign{\smallskip}
 2s22p5   &(2Po)& 3/2& 3p&  1& $-$36.7729 &  2.80 &  1~ P e\\
 2s22p5   &(2Po)& 3/2& 3p&  2& $-$35.8059 &  2.84 &  1~ D e\\
 2s22p5   &(2Po)& 1/2& 3p&  0& $-$34.8040 &  2.84 &  1~ S e\\
 \noalign{\smallskip}
 \multicolumn{8}{l}{Nlv(c)=~ 3~ : set complete
                                                                      } \\
 \noalign{\smallskip}
 \hline
\end{tabular}
\end{table}
\begin{table}
\noindent{Table 2b. Calculated \Fe17\ fine structure levels, table not extended
to symmetries other than $J\pi=0^\re$. This symmetry has {\tt Nlv = 20} levels 
below $\nu=11$ for the core ground state series: 3 Rydberg series ($\nu$
measured from the respective series limits, $E$ from the core ground state
$^2$P$_{3/2}$, the first limit).\\[.3ex]
}
\begin{tabular}{rl@{\hspace{.5em}}r@{\hspace{.4em}}rrc@{\hspace{.6em}}rr@{}}
\hline
\noalign{\smallskip}
{\tt I} &\multicolumn{3}{c}{{\it level}} &$J$
 &$E$/Ry &\multicolumn{2}{r}{$\nu\quad \ SL\pi$}\\[.5ex]
\hline\noalign{\smallskip}
\multicolumn{8}{c}{\tt Nlv= 20,\qquad J pi =  0 e}\\[.5ex]
\hline
\noalign{\smallskip}
 1 & 2s2p6  &         &       &   0      &$-$9.28398E+1 &       &$^1$S$^\re$\\
 2 & 2s22p5 &($^2$P$^\ro_{3/2}$) & 3p &0 &$-$3.62221E+1 & 2.825 &$^3$P$^\re$\\
 3 & 2s22p5 &($^2$P$^\ro_{1/2}$) & 3p &0 &$-$3.48040E+1 & 2.844 &$^1$S$^\re$\\
 4 & 2s2p6  &($^2$S$    _{1/2}$) & 3s &0 &$-$2.90350E+1 & 2.731 &$^1$S$^\re$\\
 5 & 2s22p5 &($^2$P$^\ro_{3/2}$) & 4p &0 &$-$1.95296E+1 & 3.847 &$^3$P$^\re$\\
 6 & 2s22p5 &($^2$P$^\ro_{1/2}$) & 4p &0 &$-$1.87056E+1 & 3.836 &$^1$S$^\re$\\
 7 & 2s22p5 &($^2$P$^\ro_{3/2}$) & 5p &0 &$-$1.22822E+1 & 4.850 &$^3$P$^\re$\\
 8 & 2s22p5 &($^2$P$^\ro_{1/2}$) & 5p &0 &$-$1.14454E+1 & 4.830 &$^1$S$^\re$\\
 9 & 2s2p6  &($^2$S$    _{1/2}$) & 4s &0 &$-$1.10221E+1 & 3.734 &$^1$S$^\re$\\
10 & 2s22p5 &($^2$P$^\ro_{3/2}$) & 6p &0 &$-$8.44845E+0 & 5.849 &$^3$P$^\re$\\
11 & 2s22p5 &($^2$P$^\ro_{1/2}$) & 6p &0 &$-$7.57469E+0 & 5.828 &$^1$S$^\re$\\
12 & 2s22p5 &($^2$P$^\ro_{3/2}$) & 7p &0 &$-$6.15891E+0 & 6.850 &$^3$P$^\re$\\
13 & 2s22p5 &($^2$P$^\ro_{1/2}$) & 7p &0 &$-$5.26390E+0 & 6.828 &$^1$S$^\re$\\
14 & 2s22p5 &($^2$P$^\ro_{3/1}$) & 8p &0 &$-$4.68712E+0 & 7.852 &$^3$P$^\re$\\
15 & 2s22p5 &($^2$P$^\ro_{1/2}$) & 8p &0 &$-$3.78258E+0 & 7.827 &$^1$S$^\re$\\
16 & 2s22p5 &($^2$P$^\ro_{3/2}$) & 9p &0 &$-$3.68406E+0 & 8.857 &$^3$P$^\re$\\
17 & 2s2p6  &($^2$S$    _{1/2}$) & 5s &0 &$-$3.19987E+0 & 4.733 &$^1$S$^\re$\\
18 & 2s22p5 &($^2$P$^\ro_{3/2}$) &10p &0 &$-$2.97673E+0 & 9.853 &$^3$P$^\re$\\
19 & 2s22p5 &($^2$P$^\ro_{1/2}$) & 9p &0 &$-$2.76993E+0 & 8.829 &$^1$S$^\re$\\
20 & 2s22p5 &($^2$P$^\ro_{3/2}$) &11p &0 &$-$2.45262E+0 &10.855 &$^3$P$^\re$\\
 \noalign{\smallskip}
 \hline
\end{tabular}
\end{table}

Format (ii) keeps the fine structure levels together as they emerge in the
computational procedure: for a given symmetry $J\pi$ and in energy order as
shown for $0^\re$ in table 2b, which adds up to {\tt Nlv} = 20 levels, after
the self-explanatory header line. This format should be more convenient
for easy implementation in astrophysical or other plasma modeling codes
requiring large numbers of energy levels and associated transitions. 
Here of course we have a set small and transparent enough for assignment 
by hand rather than by the new code (note how different spin-orbit 
strength is reflected in
the small difference between the quantum defects $\mu_{\rm p}$ of the two
series ---  here we are facing merely p$_{3/2}$ with $t$=1 and p$_{1/2}$
with $t$=2 because of $J$=0). The levels are identified by core configuration
$C_t$ and level $(SLJ)_t$, the outer electron quantum number $nl$, total $J$,
energy against the ionization threshold $t$=1, effective quantum number $\nu$
associated with the respective series limit $t$, and a term designation.

\subsection{Oscillator strengths for E1 transitions}

The 490 bound fine structure energy levels of \Fe17\ give rise to
26,222 dipole allowed and intercombination E1 trans\-itions. The
complete set, available electronically, contains calculated transition 
probabilities $A$, oscillator strengths $f$, and line strengths $S$ and the
level energies.

A sample subset of transitions, generated by code "stgbb", is presented 
in Table 3a. The first record of the raw {\tt stgbb} output file 
{\tt FVALUE} specifies the nuclear charge number $Z=26$, $N$=9 electrons 
in the core ion Fe{\sc~xviii}, and processing directives (e.\,g.\ 0 -- 
{\em pert\/}urbative channel coupling between {\tt RA}
and $\infty$ disabled, 1 -- {\em But\/}tle correction activated). The next two
records, headers for the subsequent \Fe17\ transition array data, identify this
array as a pair ($\emptyset\,2J_1\pi_1,\ \emptyset\,2J_2\pi_2$) of symmetries
($\pi$=0 for even and =1 for odd parity), here the electric dipole transition
$J_1=0^\re- J_2=1^\ro$. {\tt stgb} had computed $N_{J_i}=$\,20 levels of the
first symmetry (decoded in Table 2b), $N_{J_k}=$\,47 to the second, hence
$20\times 47$ subsequent records, each prefaced by a pair {\tt Ii} and {\tt Ik}
of level indices (in energy order for the respective symmetry). Their
bound state energies $E_i$ and $E_k$ below the Fe{\sc~xviii} ground
state are shown in columns 3 and 4 in reduced units $z^2$\,Ry.
The radiative result in the last three columns are the $gf$-values of the
transition $\big($see Eq.\,(\ref{eq:fij})$\big)$ in length and velocity form
and the coefficient $A$ for spontaneous emission $\big($derived in the length
form, see Eq.\,(\ref{eq:Aji})$\big)$. The signs of $gf$ are in accord
with Eq.\,(\ref{eq:fij}) and would reverse on swapping the order of symmetries
$J\pi$. Complete spectroscopic identification of the transitions can be
deduced from tables of type 2b. 
For the largest listed value, 2.301$\cdot 10^{13}$/s at {\tt(Ii,Ik = 1,5)} and
associated with excitation energy 60.846\,Ry, Table 2b verifies the initial
level as the \Fe17\ ground state; we have not presented
the odd-parity $J$=1 section but can identify {\tt Ii=5} as a low lying
state from tables 1 or 6 as 2s$^2$2p$^5$3d\,$^1$P$_1^\ro$; this transition
reappears in Table 5 with energy-adjusted 2.28(13)/s.

Table 3b, dealing with the same transition array but taken from standard
{\tt STGBB} output ("stgbb.out") that provides more physics information 
of the transitions, makes interesting reading about the internal workings of
the $R$-matrix method. While the radial wave solutions for principal quantum
number values 2 or 3 lie entirely inside the R-matrix sphere with 
radius {\tt RA}, they
have most nodes outside at values $n\approx 10$. The composition of the
dipole transition amplitude {\tt D} (before normalization) therefore changes
from dominant interior contributions {\tt DI} to large outside portions
{\tt DA} as $n$ and $n'$ increase. Perturbatively computed coupling
contributions {\tt DP} between the propagation range for {\tt DA} and infinity
equally increase, to stay only just small enough at $n=11$ to be neglected as
in Table 2a ({\tt IPERT=0}) and in fact most large scale calculations (whereas
vital in collisional work!); unlike Buttle contributions {\tt DB}, which
compensate for the rigid logarithmic boundary condition at {\tt RA}, their
computation can be fairly time consuming. Especially transition (15,29) =
$(^2$P$_{1/2}$\,8p\ 0$^\re- ^2$P$_{1/2}$\,7d\ 1$^\ro)$
reveals a subtle balance among the constituents and between
the amplitudes in length and velocity formulation.

The complete table of $f$, $S$, and $A$ for the E1 transitions in
Fe XVII that will be available electronically has slightly different format 
than Table 3a to match with the similar files for other ions (e.g. for
Fe XXI, Nahar 2000). A sample is presented in Table 4. The 
top line specifies the nuclear charge (Z = 26) and number of 
electrons in the ion, $N_{elc}$ (= 10). It is followed by sets of 
oscillator strengths belonging to pairs of symmetries, 
$J_i\pi_i~-~J_k\pi_k$. Each set starts with the transitional symmetries 
expressed in the form of $2J_i~\pi_i$ and $2J_k~\pi_k$, e.g., Table 4 
presents partial transitions for the pair of symmetries, $J=o^e-J=1^o$. 
The next line provides the number of bound levels of each symmetry, 
$N_{Ji}$ and $N_{Jk}$ as in Table 3a, which is followed
by $N_{Ji}\times N_{Jk}$ number of transitions. The first two columns
are the energy level indices, $I_i$ and $I_k$, the third and the 
fourth columns are the absolute energies, $E_i$ and $E_k$, in Rydberg 
unit. The fifth column is the $gf_L$. where $f_L$ is the oscillator 
strength in length form, and $g~=~2J+1$ is the statistical weight factor of
the initial or the lower level. A negative value for $gf$ means that
$i$ is the lower level, while a positive one means that $k$ is the
lower level. Column six is the line strength (S) and the last column
is the coefficient $A_{ki}(sec^{-1})$ for spontaneous emission.

\begin{table}
\noindent{Table 4. Sample set of $f$- and $A$-values for dipole allowed
and intercombination transitions in Fe XVII in $J\pi$ order. Notation
a$\pm$b means a$\times 10^{\pm b}$. \\}
\scriptsize
\begin{tabular}{rrlllll}
\noalign{\smallskip}
\hline  
\noalign{\smallskip}
\multicolumn{2}{c}{ 26} & \multicolumn{5}{l}{ 10} \\ 
\multicolumn{7}{l}{~~~~ 0~~~~ 0~~~~~~~ 2~~~~ 1} \\
  20 &   47 & $E_i(Ry)$ & $E_j(Ry)$ & $gf_L$ & $ S  $ & $A_{ji}(sec^{-1})$ \\
\noalign{\smallskip}
   1&    1& -9.28398+1& -3.93186+1& -1.225-1&  6.866-3&  9.396+11 \\
   1&    2& -9.28398+1& -3.84325+1& -1.010-1&  5.569-3&  8.005+11 \\
   1&    3& -9.28398+1& -3.37551+1& -8.149-3&  4.138-4&  7.617+10 \\
   1&    4& -9.28398+1& -3.29952+1& -6.222-1&  3.119-2&  5.967+12 \\
   1&    5& -9.28398+1& -3.19937+1& -2.321&  1.144-1&  2.301E+13 \\
   1&    6& -9.28398+1& -2.70352+1& -3.511-2&  1.601-3&  4.070+11 \\
   1&    7& -9.28398+1& -2.67131+1& -2.843-1&  1.290-2&  3.328+12 \\
   1&    8& -9.28398+1& -2.08737+1& -2.289-2&  9.542-4&  3.175+11 \\
   1&    9& -9.28398+1& -1.99631+1& -1.761-2&  7.249-4&  2.504+11 \\
   1&   10& -9.28398+1& -1.87276+1& -3.289-3&  1.331-4&  4.837+10 \\
   1&   11& -9.28398+1& -1.84077+1& -3.601-1&  1.451-2&  5.341+12 \\
   1&   12& -9.28398+1& -1.75506+1& -3.993-1&  1.591-2&  6.059+12 \\
   1&   13& -9.28398+1& -1.29712+1& -1.004-2&  3.771-4&  1.715+11 \\
   1&   14& -9.28398+1& -1.20521+1& -1.220-2&  4.530-4&  2.133+11 \\
   1&   15& -9.28398+1& -1.19108+1& -1.138-3&  4.219-5&  1.996+10 \\
   1&   16& -9.28398+1& -1.17427+1& -1.935-1&  7.158-3&  3.407+12 \\
   1&   17& -9.28398+1& -1.08677+1& -1.488-1&  5.446-3&  2.678+12 \\
   1&   18& -9.28398+1& -1.01812+1& -1.075-2&  3.902-4&  1.967+11 \\
   1&   19& -9.28398+1& -1.00659+1& -9.202-2&  3.335-3&  1.688+12 \\
 ...&  ...&     ...     &     ...     &   ...     &     ...   &    ...     \\
\noalign{\smallskip}
\hline
\end{tabular}
\end{table}

Line strength results from BPRM are used to compute a set of transition
probabilities $A$ and $f$-values for \Fe17\ with observed energy separation
in favour of the more uncertain calculated energies, exploiting that $S$ does
not depend on level energies (the procedure is commonly employed and was first
adopted in NIST compilations). The astrophysical models also in general use
the observed transition energies for the relevant $f$ and $A$ data. They are
more appropriate for comparison or spectral diagnostics.

Coefficients $A$ and $gf$-values have been reprocessed for all
the allowed transitions ($\Delta J=0,\pm 1$) among the observed levels.
A partial set of these transitions is presented in Table 5. The set 
comprises 342 transitions of \Fe17\ (the set is also 
available electronically).  The reprocessed transitions are moreover 
ordered according to configuration $C$ and multiplet $LS$\@. This enables 
one to obtain the $f$-values for each multiplet $LS$ and check for 
completeness of the associated levels. Completeness however also depends 
on the observed set of fine structure levels since the transitions in
the set correspond only to the observed levels (NIST). The $LS$ multiplets 
serve various comparisons with other calculations and experiment where 
fine structure transitions can not be resolved. The level index, $I_i$, 
for each energy level in the table is given next to the $g$-value 
(e.\,g.\ $g_i:I_i$) for a easy pointer to the complete $f$-file.

BPRM coefficients $A$ are compared with other calculations in Table 6, and
with available NIST data. Safronova \etal (2001) obtained data of E1, E2, M1
and M2 type for transitions $2l-3l'$ of \Fe17\ using relativistic many-body
perturbation theory (MBPT). Present results agree reasonably well yet with
noticeable scatter compared to and also within (a)--(e), in particular for
the decay of level 17 (for labels see Table 7):
2s$^2$2p$^5$3d\,$^3$P$^\ro_1 - $2s$^2$2p$^6\,^1$S$_0$. Because of poorer 
consistency for intercombination transitions --- as would
happen when varying the strength of multiplet mixing --- one might go for
inclusion of all magnetic interactions among the valence electrons: after all
there are 8 of them in this sequence, while BPRM ignores magnetic 2-body
contributions (accounting only for interaction with the two closed-shell 1s
electrons). The result marked by $\ddagger$ looks encouraging --- until one
repeats the same short calculation without such terms: $8.27\cdot 10^{10}$/s
looks sobering besides the tabulated $8.89\cdot 10^{10}$/s. This way Bhatia
and Doschek's (1992) coefficient falls into place, leaving the Cornille \etal
result --- also from \sss --- the odd case out. The blanks for Cornille
et al. in the last two transitions are not incidental, since they did not
include configurations 2s2p$^63l$ which become degenerate to 
2s$^2$2p$^53l'$ in the high Z limit, according to Layzer's scaling laws 
(Layzer 1959), that it is essential to include all the configurations of 
the complex in order to correctly reproduce the terms of the Z-expansion 
of the non-relativistic energy. FS splitting of course is a different matter,
and if 2-body magnetic interaction with the closed K shell is omitted the
effective spin-orbit parameter $\zeta_{\rm 2p}= 0.620$\,Ry (0.1484\,$\cdot
Z^4$/cm) goes up to the `bare' value of 0.684\,Ry 
(or 0.1644\,$\cdot Z^4$/cm). (For the effective spin-orbit parameter 
$\zeta$ for an orbital, see Blume and Watson 1962, Eissner et al. 1974.)
So~much about a mute point of interpreting scatter. For electric
dipole transitions the BPRM code in its present state is as good as other good
approaches but readily delivering far larger data sets than anything to date.
\begin{table}
\noindent{Table 6. Comparison of BPRM calculations for decay $A^{\rm E1}(j,1)$
to the \Fe17\ ground state $C_1T_1 =$\,2s$^2$2p$^6\,^1$S$_0$
with other work\\[.7ex]}
\begin{tabular}{r@{\hspace{.9em}}l@{\hspace{.4em}}ll@{\hspace{.7em}}l@{}}
\hline
\noalign{\smallskip}
\hline
\noalign{\smallskip}
 $j$: &\ \ $C_j$ &$T_j$ 
 &&\multicolumn{1}{l}{\normalsize\ $A(s^{-1})$}\\[-.5em]
& & &\multicolumn{1}{c}{BPRM} & \multicolumn{1}{c}{others}\\
 \noalign{\smallskip}
\hline
 \noalign{\smallskip}
 3: &2s$^2$2p$^5$3s &$^1$P$_1^\ro$  &7.96(11)
 & $8.28(11)^{\rm a}, 8.01(11)^{\rm b}, 7.75(11)^{\rm c}$\\
& & & &$8.38(11)^{\rm d}, 8.30(11)^{\rm e}, 9.40(11)^\ddagger$\\
 5: &2s$^2$2p$^5$3s & $^3$P$_1^\ro$ & 9.35(11)
 & $9.76(11)^{\rm a}, 9.44(11)^{\rm b}, 9.09(11)^{\rm c}$\\
& & & &$9.63(11)^{\rm d}, 9.34(11)^{\rm e}, 8.00(11)^\ddagger$\\[.3ex]
17: &2s$^2$2p$^5$3d & $^3$P$_1^\ro$ & 7.58(10)
 & $9.19(10)^{\rm a}, 8.27(10)^{\rm b}, 7.77(10)^{\rm c}$\\
& & & &$9.42(10)^{\rm d}, 9.00(10)^{\rm e}, 8.89(10)^\ddagger$\\[.3ex]
23: &2s$^2$2p$^5$3d & $^3$D$_1^\ro$ & 5.93(12)
 & $6.33(12)^{\rm a}, 5.68(12)^{\rm b}, 5.23(12)^{\rm c}$\\
& & & &$6.01(12)^{\rm d}, 6.01(12)^{\rm e}, 5.72(12)^\ddagger$\\
27: &2s$^2$2p$^5$3d & $^1$P$_1^\ro$ & 2.28(13)
 & $2.24(13)^{\rm a}, 2.64(13)^{\rm b}, 2.44(13)^{\rm c}$\\
& & & &$2.47(13)^{\rm d}, 2.28(13)^{\rm e}, 2.52(13)^\ddagger$\\
31: &2s2p$^6$3p & $^3$P$_1^\ro$ & 4.03(11)
 & $4.51(11)^{\rm a}, 3.66(11)^{\rm b}$\\
& & & &$4.12(11)^{\rm d}, 3.40(11)^{\rm e}, 3.52(11)^\ddagger$\\
33: &2s2p$^6$3p & $^1$P$_1^\ro$ & 3.30(12)
 & $3.34(12)^{\rm a}, 3.21(12)^{\rm b}$\\
& & & &$3.29(12)^{\rm d}, 3.30(12)^{\rm e}, 3.25(12)^\ddagger$\\[.5ex]
\hline
\noalign{\smallskip}
\multicolumn{5}{l}{a~--~Safronova \etal 2001, b~--~Bhatia \& Doschek 1992,}\\
\multicolumn{5}{l}{c~--~Cornille \etal 1994, d~--~ present MCDF, e~--~ NIST,}\\
\multicolumn{5}{l}{$\ddagger$ -- \sss\ with {\em all\/} magnetic
FS-components.}\\
\noalign{\smallskip}
\hline
\end{tabular}
\end{table}

Among forbidden transitions, discussed in the next section, there is one
class for which it is obvious that one must draw very different conclusions,
that is for transitions between levels of a FS multiplet: to start with,
the splitting changes significantly on including 2-body FS contributions.

\subsection{Forbidden transitions M1, E2, and M2, E3}

We extend the behavioural study of computed radiative decay in Table 8 
to a {\em selection\/} of forbidden transitions; a {\em complete\/} set 
will be published in electronic format, available from the CDS library 
for some 3000 transitions between the 89 \Fe17-levels. Table~8 along 
with Table~7 probes the quality of the target representations --- 
especially term coupling, which is crucial in the
collisional application (CPE02). Larger uncertainties are confined to
intercombination lines, but there they can increase uncomfortably with
higher radiative multipole type. Moreover the table assesses the 
influence of 2-body finestructure contributions neglected in the current 
BPRM work. Magnetic interaction between valence shell electrons is 
always present in the MCDF work with {\small GRASP}, activated for the 
\sss\ column {\sc ss}$\sp+$ but switched
off in {\sc ss}$\sp-$: follow the trend from {\sc ss}$\sp-$
via {\sc ss}$\sp+$ to full relativistic MCDF.

At wave lengths of 10\,\AA\,$\approx$\,911\,\AA/100 (hence 
$E_{ij}^2=10^4$ Ry) Eqs (\ref{eq:E3}--\ref{eq:M2}) versus (\ref{eq:E2}--
\ref{eq:M1}) suggest a close look at decay by electric octopole and 
magnetic quadrupole radiation for transitions with such a
lowest path. We can indeed expect rates around $10^6$/s, which would be
competitive with E2 and M1 decay around Fe with $Z_{\rm eff}\approx 20$ along
the Ne-isoelectronic sequence, as the scaling laws show: inserting
(\ref{eq:Elam}) for E$\lambda$ and (\ref{eq:Mlam}) for M$\lambda$ into the line
strength expression (\ref{eq:Xlam}) yields scaling of $A$ as $Z^8$ for both
E3 and M2 (and $Z^6$ for E2 and M1); for transitions within a principal shell
($n$\,=\,0) though scaling of E$\lambda$ drops by a factor of $Z^2$, and 
octopole transitions become negligible. The E3 results in Table 8 are most
satisfactory and perfectly understood. To start with the two bottom entries,
one of them apparently contradicting this statement, Table 7 identifies levels
87 and 89 as multiplet mixing companions with $J\,=\,3$ to terms 4f $^3$F
and $^1$F\@. Therefore the intercombination decay of 87 becomes rather
sensitive to magnetic coupling, $A$ converging from right to left as much
as one can reasonably expect when MCDF works with a slightly different
target. This is borne out by 56, the only other troubling level for E3,
as Table 7 places it marginally differently (unfortunately no experiment has
yet decided). M2 is a different matter, a factor of 2.5 in the poor case
(18,1) difficult to reconcile with the lowest order radiative operator as
adopted in \sss.

For E2 vs M1 the picture turns very varied as early as for $\Delta n\neq 0$:
distinguishing between intercombination transitions (with factors
like $\alpha^2 Z^2$ and $\alpha^2 Z^3$) and direct transition becomes a more
persistent companion. For direct transitions between main shells both $A$
scale as $Z^6$, the time coefficient favouring E2. Next come radiative BP
corrections to M1 remembered from the classical case of 1s2s\,$^3$S decay.
We verified the Bhatia and Doschek entries, converting to $A$ without those
corrections with the help of an expedient tool: \sss\ prints both the full
line strength $S^{\rm M1}$ and BP-deficient $S_0^{\rm M1}$. Then $A(9,1)$
drops to less than its tenth, from its {\sc ss}$\sp+$ result 3.31$\times10^3$
sec$^{-1}$ --- albeit only half what MCDF is telling: greater discrepancies are
associated with differences between SS$\sp+$ and SS$\sp-$ results and rather
crowded fields in Table 7 for the respective $J\pi$, so BP may be stretched
beyond its limits. The trends for E2 type transitions look perfect.

For electric dipole transitions, both direct and spin-flip, Table 8 gives $A$
in velocity form as a second entry to the more firmly established length
results, as a measure of good target description $\big($with the proviso after
Eq.\,(7)$\big)$. They compare encouragingly for the EIE work.

Turning briefly towards astrophysical and laboratory implications from Table 8,
apart from selected spontaneous emission coefficients for dipole-allowed
transitions it gives results to magnetic dipole and electric quadrupole
radiation --- and some magnetic quadrupole and electric octopole transitions of
the same magnitude of some  $10^6$ s$^{-1}$: of course this high multipole
decay mode can compete only for transitions with very short wave length,
i.\,e.\ to the ground state. It may influence the modeling of line emissions.
In astronomy and in laboratory photoionized plasmas
the M2 decay from level 2 has long been observed as a prominent line.
The population of level 2 is fed by cascading from 2p$^5$3s, 2p$^5$3p,
and 2p$^5$3d and higher configurations. Accurate M2 transition
probabilities are the key to modeling this line.
Moreover it has important plasma diagnostics potential.
\begin{table}
\noindent{Table 8. Selected transition probabilities $A\cdot$\,s of
Fe{\sc~xvii}, for electric dipole E1 type transitions also in velocity
formulation as second entries, computed by \sss\ with and without
2-body FS-terms (columns {\sc ss}$\sp+$
and {\sc ss}$\sp-$) and MCDF, and miscellaneous results: \
E1 --- from BPRM, \quad\
M1 --- $A^{\rm M1}\cdot$\,s by Bhatia and Doschek (1992)
employing (\ref{eq:Mlam}) rather than full (\ref{eq:OM1}), \quad\
E2 --- from BPRM\@.
The quantity $a$e$b$ stands for $a\times10^b$.}
\begin{center}
\begin{tabular} {rrccccc}
\hline
            \noalign{\smallskip}
 $i$ & $j$ &{\it type} &{\sc mcdf} &{\sc ss}$\sp+$ & {\sc ss}$\sp-$ &misc.\\
\hline
            \noalign{\smallskip}
 3&1 &E1 &9.63e11 &9.39e11 &9.42e11 &9.39e11\\
  &&& 9.24e11 &8.43e11 &8.51e11 &9.44e11\\
 5&1 &E1 &8.38e11 &8.00e11 &7.98e11 &8.01e11\\
  &&& 8.02e11 &7.76e11 &7.73e11 &8.08e11\\
17&1 &E1 &9.42e10 &8.89e10 &8.23e10 &7.61e10\\
  &&& 8.73e10 &8.27e10 &7.65e10 &7.49e10\\
23&1 &E1&6.01e12 &5.72e12 &5.73e12 & 5.96e12\\
  &&& 5.65e12 &5.39e12 &5.41e12 &5.69e12\\
27&1 &E1 &2.47e13 &2.52e13 &2.52e13 &2.30e13\\
  &&& 2.32e13 &2.40e13& 2.41e13 &2.19e13\\
33&1 &E1&3.29e12 &3.25e12 &3.25e12 & 3.32e12\\
  &&& 3.30e12 &3.39e12&3.38e12 &3.49e12\\[1.2ex]
       6&1 &M1 &1.80e5 &1.74e5 &1.61e5&4.96e+4\\
       9&1 &M1 &6.81e3 &3.31e3 &4.43e3&5.94e+4\\
      12&1 &M1 &4.24e3 &4.98e3 &4.34e3&2.20e+3\\
      13&1 &M1 &2.03e5 &1.77e5 &1.79e5&1.99e+5\\
      28&1 &M1 &1.93e4 &1.97e4 &1.76e4&2.33e+1\\
      34&1 &M1 &2.10e3 &5.31e3 &8.25e3&1.67e$-$1\\[1.2ex]
 7&1 &E2 &5.24e08 &5.14e08 &5.16e08 &5.15e08\\
10&1 &E2 &5.63e08 &5.62e08 &5.60e08 &5.52e08\\
14&1 &E2 &6.77e08 &6.63e08 &6.62e08 &6.69e08\\
35&1 &E2 &1.86e07 &2.52e07 &4.01e07 &5.85e07\\
37&1 &E2 &1.09e10 &1.08e10 &1.08e10 &1.10e10\\
85&1 &E2 &3.00e09 &2.98e09 &2.98e09
\\[1.2ex]
       2&1 &M2 &2.25e5 &2.17e5 &2.17e5&\\ 
      18&1 &M2 &6.16e6 &2.58e6 &2.63e6&\\
      21&1 &M2 &1.13e6 &6.27e5 &5.44e5&\\
      24&1 &M2 &4.47e5 &8.28e5 &8.79e5&\\
      25&1 &M2 &2.73e5 &4.15e6 &4.14e6&\\
      32&1 &M2 &8.44e5 &8.02e5 &8.02e5&\\[1.2ex]
20&1 &E3 &2.83e5 &2.82e5 &2.85e5\\
22&1 &E3 &3.52e5 &3.61e5 &3.60e5\\
26&1 &E3 &4.00e5 &3.94e5 &3.93e5\\
56&1 &E3 &3.87e4 &1.48e5 &1.49e5\\
87&1 &E3 &1.23e5 &1.92e5 &2.75e5\\
89&1 &E3 &3.36e6 &3.64e6 &3.56e6\\
\noalign{\smallskip}
\hline
\end{tabular}
\end{center}
\end{table}

\section{Conclusions}

>From large-scale state-of-the-art calculations in Breit-Pauli approximation
we obtain energy levels with principal quantum number up to $\,n=\,10$ and
radiative transition probabilities of \Fe17. All levels have been identified
in spectroscopic notation and checked for completeness. The set of results 
far exceeds the currently available experimental and theoretical data. 

Radiative data for most electric dipole transions as well as level positions
agree within 10\% and in most cases far better with available theoretical and
experimental work of quality. This indicates that for these highly charged
ions higher order relativistic and QED effects omitted in the BPRM
calculations may lead to an error not exceeding the estimated uncertainty. 

We have obtained a consistent set of coefficients $A$ for E2 and M1 type
transitions and compared our \sss\ and MCDF calculations and with other
calculations in the literature. Most results for $A^{\rm E2}$ and $A^{\rm M1}$
lie well inside 20--30\% of uncertainty. However, numerically very small
coefficients can differ from 50\% to a factor of two: M2 and in particular
E3 results are highly sensitive to the physics included and numerics (e.\,g.\
cancellation effects and numerical instabilities). Large differences are found
between the \sss\ and MCDF calculations. Especially the magnetic quadrupole
results are hard to assess, suggesting further study of this issue.

All data are available electronically. Part of the $f$-values have been
reprocessed using available observed energies for better accuracy.
The new results should be particularly useful for
the analysis of X-ray and Extreme Ultraviolet spectra from 
astrophysical and laboratory sources where non-local thermodynamic 
equilibrium (NLTE) atomic models with many excited levels are needed.
%
\begin{acknowledgements}
This work was partially supported by U.S. National Science Foundation 
(AST-9870089) and the NASA ADP program; WE enjoyed part-support by
Sonderforschungsbereich 392 of the German Research Council.
The computational work was largely carried out on the Cray T94 and
Cray SV1 at the Ohio Supercomputer Center in Columbus, Ohio.
\end{acknowledgements}

%

\onecolumn
\begin{table*}
\noindent{Table 3a. Truncated {\tt STGBB} output `{\tt FVALUE}':
$gf$-values and Einstein coefficients $A$ for [\verb|0 0 0   0 2 1|]
$=\,(0^\re - 1^\ro)$ transitions of \Fe17\ [{\tt Z=26, \mbox{\rm core-}Nel=9}],
as function of bound state energies {\tt RE}($n_1l_1\,0^\re$)
and {\tt Re}($n_2l_2\,1^\ro$}) in units of $z^2\,$Ry, $z$\,={\tt\,26-9}.
The line strength column S(E1) has been added by hand
$\big($see Eqs (7--8)$\big)$.
\begin{verbatim}
        26         9    IPERT= 0     AC,IBUT= 1.0E-5 1  06/25/01  15:06:37
    0    0    0         0    2    1
   20   47    RE1          RE2     GFL  -  E1  -  GFV    A(E1)*s      S(E1) 
  1  1 -3.212451E-1 -1.360506E-1 -1.225E-01 -1.232E-01  9.396E+11   6.866E-3
  1  2 -3.212451E-1 -1.329843E-1 -1.010E-01 -1.020E-01  8.005E+11   5.569E-3
  1  3 -3.212451E-1 -1.167997E-1 -8.149E-03 -8.015E-03  7.617E+10   4.138E-4
  1  4 -3.212451E-1 -1.141704E-1 -6.222E-01 -5.940E-01  5.967E+12   3.119E-2
  1  5 -3.212451E-1 -1.107050E-1 -2.321E+00 -2.214E+00  2.301E+13   1.144E-1
  1  6 -3.212451E-1 -9.354728E-2 -3.511E-02 -3.404E-02  4.070E+11   1.601E-3
  1  7 -3.212451E-1 -9.243291E-2 -2.843E-01 -2.989E-01  3.328E+12   1.290E-2
  1  8 -3.212451E-1 -7.222739E-2 -2.289E-02 -2.328E-02  3.175E+11   9.542E-4
  1  9 -3.212451E-1 -6.907663E-2 -1.761E-02 -1.707E-02  2.504E+11   7.249E-4
  1 10 -3.212451E-1 -6.480154E-2 -3.289E-03 -3.310E-03  4.837E+10   1.331E-4
  1 11 -3.212451E-1 -6.369437E-2 -3.601E-01 -3.275E-01  5.341E+12   1.451E-2
  1 12 -3.212451E-1 -6.072860E-2 -3.993E-01 -3.613E-01  6.059E+12   1.591E-2
  1 13 -3.212451E-1 -4.488317E-2 -1.004E-02 -1.027E-02  1.715E+11   3.771E-4
  1 14 -3.212451E-1 -4.170278E-2 -1.220E-02 -1.168E-02  2.133E+11   4.530E-4
  1 15 -3.212451E-1 -4.121385E-2 -1.138E-03 -1.113E-03  1.996E+10   4.219E-5
  1 16 -3.212451E-1 -4.063208E-2 -1.935E-01 -1.755E-01  3.407E+12   7.158E-3
  1 17 -3.212451E-1 -3.760452E-2 -1.488E-01 -1.349E-01  2.678E+12   5.446E-3
  1 18 -3.212451E-1 -3.522922E-2 -1.075E-02 -1.194E-02  1.967E+11   3.902E-4
  1 19 -3.212451E-1 -3.483027E-2 -9.202E-02 -9.137E-02  1.688E+12   3.335E-3
 .....
 20 45 -8.486569E-3 -8.686647E-3  7.494E-05  5.519E-05  2.012E+03   3.822E-3
 20 46 -8.486569E-3 -8.368791E-3 -4.901E+00 -4.929E+00  1.520E+07   4.319E+2
 20 47 -8.486569E-3 -8.321573E-3 -3.850E-01 -3.826E-01  2.344E+06   2.422E+1
    0    2    0         0    0    1
   45   19    RE1          RE2     GFL  -  E1  -  GFV    A(E1)*S
  1  1 -1.288567E-1 -1.332897E-1  2.198E-03  9.246E-04  9.659E+06
  1  2 -1.288567E-1 -1.170725E-1 -1.226E-01 -1.213E-01  1.142E+10
  1  3 -1.288567E-1 -9.365835E-2 -5.438E-02 -3.476E-02  4.520E+10
 .....
    0    0    0         0    0    0
 _________________________________________________________________________
\end{verbatim}
\end{table*}

\begin{table*}
\noindent{Table 3b. Truncated {\tt STGBB} standard output: array
$\,(0^\re - 1^\ro)$ of \Fe17, build-up of\\ the dipole transition amplitude
{\tt D} by the {\it R-}matrix code $\big($L[ength] and V[elocity]$\big)$.}
\begin{verbatim}
  IPERT = 1    AC = 1.00E-05:       BOUND-BOUND TRANSITION DATA FOR
               R0 = 40.3750

          (IS1,IL1,IP1) = (  0  0  0 )     (IS2,IL2,IP2) = (  0  2  1 )

  I  J TYPE   DI         DA         DB         DP        D          S

  1  1  L  1.409E+00 -2.690E-07  2.374E-07 -3.00E-10  1.409E+00  6.867E-03
        V  1.413E+00  2.111E-08 -3.675E-08 -4.19E-10  1.413E+00  6.905E-03

  1  2  L -1.269E+00  1.270E-06 -2.396E-07  3.27E-10 -1.269E+00  5.569E-03
        V -1.275E+00 -1.017E-07  3.502E-08  4.08E-10 -1.275E+00  5.623E-03

  1  3  L  3.458E-01  1.759E-07 -5.994E-09 -3.39E-11  3.458E-01  4.138E-04
        V  3.429E-01 -9.418E-09  1.170E-09 -2.36E-11  3.429E-01  4.070E-04

 .....
  1 47  L -3.690E-01 -2.725E-04  4.084E-07 -1.58E-06 -3.693E-01  4.718E-04
        V -3.498E-01  1.021E-06  3.940E-07 -3.67E-08 -3.498E-01  4.235E-04

  2  1  L -5.324E+00 -3.248E-06  8.932E-09 -2.74E-09 -5.324E+00  9.807E-02
        V -5.246E+00 -3.527E-06 -2.929E-09 -4.12E-09 -5.246E+00  9.522E-02

  2  2  L -3.360E+00  2.728E-07  2.571E-09  1.10E-08 -3.360E+00  3.906E-02
        V -3.157E+00 -9.540E-07 -3.769E-08  1.04E-08 -3.157E+00  3.448E-02
 .....
 15 29  L -1.200E+00  1.125E+01  9.775E-05 -2.27E-03  1.005E+01  3.495E-01
        V  1.501E+00  8.534E+00 -9.112E-04 -1.70E-03  1.003E+01  3.483E-01
 .....
 20 47  L  5.763E-01  8.471E+01 -3.862E-04 -2.22E+00  8.307E+01  2.388E+01
        V  2.870E+00  8.220E+01 -4.878E-02 -2.21E+00  8.282E+01  2.373E+01
 _________________________________________________________________________
\end{verbatim}
\end{table*}


\begin{table*}
\noindent{Table 5. Dipole allowed and intercombination transitions in \Fe17.
The calculated transition energies are replaced by {\em observed\/} energies.
The $g:I$ indices refer to the statistical weight:energy level index in the
raw data file. The notation $a(b)$ means $a\times 10^b$.\\[1.ex]}
\begin{tabular}{llllllrcc}
\noalign{\smallskip}
\hline
\noalign{\smallskip}
$C_i$ & $C_j$  & $T_i$ & $T_j$ & $g_i:I_i$ & $g_j:I_j$ & $\lambda_{ij}$/\AA &
$f$ & $A\cdot$\,s\\
 \noalign{\smallskip}
\hline
 \noalign{\smallskip}
 2p6     & 2s22p53s  & $^1$S$^\re$ & $^3$P$^\ro$ &  1: 1 &  3: 1 &  17.1 & 1.223($-$1) &  9.35(11)\\[.8ex]
 2p6     & 2s22p63s  & $^1$S$^\re$ & $^1$P$^\ro$ &  1: 1 &  3: 2 &  16.8 & 1.008($-$1) &  7.96(11)\\[.8ex]
 2p6     & 2s22p53d  & $^1$S$^\re$ & $^3$P$^\ro$ &  1: 1 &  3: 3 &  15.4 & 8.136($-$3) &  7.58(10)\\[.8ex]
 2p6     & 2s22p53d  & $^1$S$^\re$ & $^3$D$^\ro$ &  1: 1 &  3: 4 &  15.3 & 6.208($-$1) &  5.93(12)\\[.8ex]
 2p6     & 2s22p53d  & $^1$S$^\re$ & $^1$P$^\ro$ &  1: 1 &  3: 5 &  15.0 & 2.314 &  2.28(13)\\[.8ex]
 2p6     & 2s2p63p   & $^1$S$^\re$ & $^3$P$^\ro$ &  1: 1 &  3: 6 &  13.9 & 3.501($-$2) &  4.03(11)\\[.8ex]
 2p6     & 2s2p63p   & $^1$S$^\re$ & $^1$P$^\ro$ &  1: 1 &  3: 7 &  13.8 & 2.835($-$1) &  3.30(12)\\[.8ex]
 2p6     & 2s22p54s  & $^1$S$^\re$ & $^3$P$^\ro$ &  1: 1 &  3: 8 &  12.7 &  2.286($-$2) &  3.16(11)\\[.8ex]
 2p6     & 2s22p54s  & $^1$S$^\re$ & $^1$P$^\ro$ &  1: 1 &  3: 9 &  12.5 &  1.758($-$2) &  2.49(11)\\[.8ex]
 2p6     & 2s22p54d  & $^1$S$^\re$ & $^3$P$^\ro$ &  1: 1 &  3:10 &  12.3 &  3.281($-$3) &  4.81(10)\\[.8ex]
 2p6     & 2s22p54d  & $^1$S$^\re$ & $^3$D$^\ro$ &  1: 1 &  3:11 &  12.3 &  3.594($-$1) &  5.31(12)\\[.8ex]
 2p6     & 2s22p54d  & $^1$S$^\re$ & $^1$P$^\ro$ &  1: 1 &  3:12 &  12.1 &  3.987($-$1) &  6.03(12)\\[.8ex]
 2p6     & 2s22p55s  & $^1$S$^\re$ & $^3$P$^\ro$ &  1: 1 &  3:13 &  11.4 &  1.003($-$2) &  1.71(11)\\[.8ex]
 2p6     & 2s22p55s  & $^1$S$^\re$ & $^1$P$^\ro$ &  1: 1 &  3:14 &  11.3 &  1.219($-$2) &  2.13(11)\\[.8ex]
 2p6     & 2s22p55d  & $^1$S$^\re$ & $^3$P$^\ro$ &  1: 1 &  3:15 &  11.3 &  1.135($-$3) &  1.98(10)\\[.8ex]
 2p6     & 2s22p55d  & $^1$S$^\re$ & $^3$D$^\ro$ &  1: 1 &  3:16 &  11.3 &  1.932($-$1) &  3.39(12)\\[.8ex]
 2p6     & 2s22p55d  & $^1$S$^\re$ & $^1$P$^\ro$ &  1: 1 &  3:17 &  11.1 &  1.486($-$1) &  2.67(12)\\[.8ex]
 2p6     & 2s2p64p   & $^1$S$^\re$ & $^3$P$^\ro$ &  1: 1 &  3:18 &  11.0 &  1.073($-$2) &  1.96(11)\\[.8ex]
 2p6     & 2s2p64p   & $^1$S$^\re$ & $^1$P$^\ro$ &  1: 1 &  3:19 &  11.0 &  9.190($-$2) &  1.68(12)\\[.8ex]
 2p53s   & 2s22p53p  & $^3$P$^\ro$ & $^3$P$^\re$ &  3: 1 &  1: 2 & 296.0 &  3.354($-$2) &  7.66(09)\\
 2p53s   & 2s22p53p  & $^3$P$^\ro$ & $^3$P$^\re$ &  3: 1 &  3: 4 &    262.7 &  5.893($-$5) &  5.70(06)\\
 2p53s   & 2s22p53p  & $^3$P$^\ro$ & $^3$P$^\re$ &  5: 1 &  3: 4 & 252.5 &  4.985($-$3) &  8.69(08)\\
 2p53s   & 2s22p53p  & $^3$P$^\ro$ & $^3$P$^\re$ &  3: 1 &  5: 2 & 340.4 &  9.075($-$2) &  3.13(09)\\
 2p53s   & 2s22p53p  & $^3$P$^\ro$ & $^3$P$^\re$ &  5: 1 &  5: 2 & 323.5 &  6.913($-$2) &  4.41(09)\\
 \multicolumn{2}{l}{$LS$} & $^3$P$^\ro$ & $^3$P$^\re$ &  9 &  9 &  &  8.959($-$2) & 6.71(09) \\
 & & & & & & & &\\
 2p63s   & 2s22p53p  & $^1$P$^\ro$ & $^3$P$^\re$ &  3: 2 &  1: 2 & 413.8 &  9.557($-$3) &  1.12(09)\\
 2p63s   & 2s22p53p  & $^1$P$^\ro$ & $^3$P$^\re$ &  3: 2 &  3: 4 & 351.6 &  4.162($-$2) &  2.25(09) \\
 2p63s   & 2s22p53p  & $^1$P$^\ro$ & $^3$P$^\re$ &  3: 2 &  5: 2 & 506.3 &  1.464($-$3) &  2.29(07)\\
 & & & & & & & & \\
 2p53p   & 2s22p53d  & $^3$P$^\re$ & $^3$P$^\ro$ &  1: 2 &  3: 3 & 369.5 &  9.560($-$3) &  1.56(08)\\
 2p53p   & 2s22p53d  & $^3$P$^\re$ & $^3$P$^\ro$ &  3: 4 &  1: 2 & 457.5 &  1.443($-$3) &  1.38(08)\\
 2p53p   & 2s22p53d  & $^3$P$^\re$ & $^3$P$^\ro$ &  3: 4 &  3: 3 & 439.0 &  1.262($-$3) &  4.37(07)\\
 2p53p   & 2s22p53d  & $^3$P$^\re$ & $^3$P$^\ro$ &  3: 4 &  5: 2 & 415.7 &  5.280($-$4) &  1.22(07)\\
 2p53p   & 2s22p53d  & $^3$P$^\re$ & $^3$P$^\ro$ &  5: 2 &  3: 3 & 317.7 &  9.904($-$3) &  1.09(09)\\
 2p53p   & 2s22p53d  & $^3$P$^\re$ & $^3$P$^\ro$ &  5: 2 &  5: 2 & 305.4 & 5.093($-$2) &  3.64(09)\\
 \multicolumn{2}{l}{$LS$} & $^3$P$^\re$ & $^3$P$^\ro$ &  9 &  9 &  & 3.145($-$2) &  1.64(09) \\
 & & & & & & & &\\
 2p53p   & 2s22p53d   & $^3$P$^\re$ & $^3$D$^\ro$ &  1: 2 &  3: 4 &    285.5 &  2.019($-$1) &  5.51(09)\\
 2p53p   & 2s22p53d   & $^3$P$^\re$ & $^3$D$^\ro$ &  3: 4 &  3: 4 &    325.2 &  2.756($-$3) &  1.74(08)\\
 2p53p   & 2s22p53d   & $^3$P$^\re$ & $^3$D$^\ro$ &  3: 4 &  5: 5 & 279.9 &  1.945($-$1) &  9.93(09)\\
 2p53p   & 2s22p53d   & $^3$P$^\re$ & $^3$D$^\ro$ &  5: 2 &  3: 4 & 253.6 &  1.599($-$5) &  2.76(06)\\
 2p53p   & 2s22p53d   & $^3$P$^\re$ & $^3$D$^\ro$ &  5: 2 &  5: 5 & 225.1 &  4.933($-$3) &  6.49(08)\\
 2p53p   & 2s22p53d   & $^3$P$^\re$ & $^3$D$^\ro$ &  5: 2 &  7: 2 &  280.1 &  1.573($-$1) &  9.55(09)\\
 \multicolumn{2}{l}{$LS$} & $^3$P$^\re$ & $^3$D$^\ro$ &9 &$\!\!15$&& 1.887($-$1) &  1.08(10)\\
 & & & & & & & &\\
 2p53p  & 2s22p53d   & $^3$P$^\re$ & $^1$P$^\ro$ &  1: 2 &  3: 5 &    218.3 &  1.528($-$2) &  7.13(08)\\
 2p53p  & 2s22p53d   & $^3$P$^\re$ & $^1$P$^\ro$ &  3: 4 &  3: 5 &    240.8 &  2.555($-$2) &  2.94(09)\\
 2p53p  & 2s22p53d   & $^3$P$^\re$ & $^1$P$^\ro$ &  5: 2 &  3: 5 &    199.1 &  1.793($-$4) &  5.02(07)\\
 & & & & & & & & \\
 2p53p  & 2s2p63p  & $^3$P$^\re$ & $^3$P$^\ro$ &  1: 2 &  3: 6 & 100.3 &  3.128($-$2) &  6.91(09)\\
 2p53p  & 2s2p63p  & $^3$P$^\re$ & $^3$P$^\ro$ &  3: 4 &  3: 6 & 104.8 &  2.500($-$3) &  1.52(09)\\
 2p53p  & 2s2p63p  & $^3$P$^\re$ & $^3$P$^\ro$ &  3: 4 &  3: 6 & 104.8 &  2.500($-$3) &  1.52(09)\\
 2p53p  & 2s2p63p  & $^3$P$^\re$ & $^3$P$^\ro$ &  5: 2 &  3: 6 &  96.1 &  2.663($-$3) &  3.21(09)\\
\multicolumn{2}{l}{$LS$} & $^3$P$^\re$ & $^3$P$^\ro$ &  9 &  9 &  & 5.822($-$3) &  3.94(09)\\
 \noalign{\smallskip}
\hline
\end{tabular}
\end{table*}

\newpage

\begin{table*}
{Table 7. The first 89 fine-structure $n=2$, 3 and 4 levels included in 
the EIE calculation by Chen \etal 2003: comparison of calculated and
observed energies in Rydbergs for Fe {\sc xvii}; `obs' data are observed
values from NIST; the entries `{\sc ss}' ({\sc ss}$\sp-$/{\sc ss}$\sp+$:
without/with inclusion of 2-body magnetic components) and the entries
`{\sc mcdf}' are from \sss\ and {\sc grasp} calculations respectively.\\[1.ex]}
\begin{tabular} {rrrlrcrc}
\hline
\noalign{\smallskip}
 $i$ &{\it SLJ} &{\it (jj)\,J} &\ \ obs &{\sc ss}$\sp-$ & ss$\sp+$
 &{\sc mcdf} & BPRM\\
\noalign{\smallskip}
\hline
\noalign{\smallskip}
  1&2s$^2$2p$^{6\ ^1}$S$_0$ &(0,0)0&0.0&0.0&0.0&0.0 &0.0\\
  2&2s$^2$2p$^5$3s\ $^3$P$_2^\ro$ &(3/2,1/2)$^\ro$2&53.2965&53.3622 &53.3666
 &53.1684&53.3821\\
  3&   3s\ $^1$P$_1^\ro$ &(3/2,1/2)$^\ro$1&53.43&53.5044 &53.5091 &53.3100
 &53.5211\\
  4&   3s\ $^3$P$_0^\ro$ &(1/2,1/2)$^\ro$0&54.2268&54.2865 &54.2865 &54.0957
 &54.3190\\
  5&   3s\ $^3$P$_1^\ro$ &(1/2,1/2)$^\ro$1&54.3139&54.3791 &54.3697 &54.1851
 &54.4074\\[.7ex]
  6&   3p\ $^3$S$_1$ &(3/2,1/2)1&55.5217&55.5686 &55.5735 &55.3963&55.6001\\
  7&   3p\ $^3$D$_2$ &(3/2,1/2)2&55.7787&55.8397 &55.8455 &55.6606&55.8654\\
  8&   3p\ $^3$D$_3$ &(3/2,3/2)3&55.8974&55.9463 &55.9494 &55.7791&55.9857\\
  9&   3p\ $^1$P$_1$ &(3/2,3/2)1&55.9804&56.0338 &56.0404 &55.8654&56.7674\\
 10&   3p\ $^3$P$_2$ &(3/2,3/2)2&56.1137&56.1597 &56.1642 &55.9950&56.2007\\
 11&   3p\ $^3$P$_0$ &(3/2,3/2)0&56.5155&56.5821 &56.5809 &56.4050&56.2221\\
 12&   3p\ $^3$D$_1$ &(1/2,1/2)1&56.6672&56.7288 &56.7211 &56.5495&56.0669\\
 13&   3p\ $^3$P$_1$ &(1/2,3/2)1&56.9060&56.9499 &56.9420 &56.7855&57.0024\\
 14&   3p\ $^1$D$_2$ &(1/2,3/2)2&56.9336&56.9817 &56.9703 &56.8135&57.0339\\
 15& 3p\ $^1$S$_0$ &(1/2,1/2)0&57.8894&58.0639 &58.0619 &57.9308&58.0358\\[.7ex]
 16&   3d\ $^3$P$_0^\ro$ &(3/2,3/2)$^\ro$0&58.8982&58.9407 &58.9578 &58.7738&59.0057\\
 17&   3d\ $^3$P$_1^\ro$ &(3/2,3/2)$^\ro$1&58.981&59.0188 & 59.0289 &58.8454&59.0846\\
 18&   3d\ $^3$P$_2^\ro$ &(3/2,5/2)$^\ro$2&59.0976&59.1651 &59.1659 &58.9826&59.2305\\
 19&   3d\ $^3$F$_4^\ro$ &(3/2,5/2)$^\ro$4&59.1041&59.1821 &59.1799 &58.9901&59.2435\\
 20&   3d\ $^3$F$_3^\ro$ &(3/2,3/2)$^\ro$3&59.1611&59.2240 &59.2347 &59.0498&59.2820\\
 21&   3d\ $^1$D$_2^\ro$ &(3/2,3/2)$^\ro$2&59.2875&59.3513 &59.3630 &59.1797&59.4106\\
 22&   3d\ $^3$D$_3^\ro$ &(3/2,5/2)$^\ro$3&59.3665&59.4471 &59.4466 &59.2598
 &59.5054\\
 23&   3d\ $^3$D$_1^\ro$ &(3/2,5/2)$^\ro$1&59.708&59.7865 & 59.7907 &59.6082
 &59.8446\\
 24&   3d\ $^3$F$_2^\ro$ &(1/2,3/2)$^\ro$2&60.0876&60.1438 &60.1431 &59.9749
 &60.2171\\
 25&   3d\ $^3$D$_2^\ro$ &(1/2,5/2)$^\ro$2&60.1617&60.2179 &60.2045 &60.0344
 &60.2940\\
 26&   3d\ $^1$F$_3^\ro$ &(1/2,5/2)$^\ro$3&60.197&60.2627 & 60.2484 &60.0754&60.3337\\
 27&   3d\ $^1$P$_1^\ro$ &(1/2,3/2)$^\ro$1&60.6903&60.8225 &60.8212 &60.6279&60.8461\\[.9ex]
 28&2s2p$^6$3s\ $^3$S$_1$ &(1/2,1/2)1&&63.3306 &63.3306 &63.2125 &63.3658\\
 29&   3s\ $^1$S$_0$ &(1/2,1/2)0&&63.7925 &63.7925 &63.6986 &63.8049\\
 30&   3p\ $^3$P$_0^\ro$ &(1/2,1/2)$^\ro$0&&65.7338&65.7377&65.6346 &65.7726\\
 31&   3p\ $^3$P$_1^\ro$ &(1/2,1/2)$^\ro$1 &65.601 &65.7687 &65.7703&65.6676
 &65.8047\\
 32&   3p\ $^3$P$_2^\ro$ &(1/2,3/2)$^\ro$2&&65.9299&65.9285&65.8380 &65.9792\\
 33&   3p\ $^1$P$_1^\ro$ &(1/2,3/2)$^\ro$1&65.923&66.0723 &66.0718 &65.9782
 &66.1267\\
 34&   3d\ $^3$D$_1$ &(1/2,3/2)1& &69.0162 &69.0269&68.9221 &69.0744\\
 35&   3d\ $^3$D$_2$ &(1/2,3/2)2& &69.0351 &69.0386&68.9323 &69.0920\\
 36&   3d\ $^3$D$_3$ &(1/2,5/2)3& &69.0672 &69.0606&68.9518 &69.1237\\
 37&   3d\ $^1$D$_2$ &(1/2,5/2)2&69.282&69.4358&69.4352&69.3247 &69.4813\\[.7ex]
 38&2s$^2$2p$^5$4s\ $^3$P$_2^\ro$ && &71.8710 &71.8754 &71.6517\\
 39&2s$^2$2p$^5$4s\ $^1$P$_2^\ro$ &&71.860 &71.9150 &71.9197 &71.6983\\[.9ex]
\vdots\\
 55&                 $^3$P$_2^\ro$ &             &&74.0927 &74.1062 &73.9033\\
 56&2s$^2$2p$^5$4d\ $^3$F$_3^\ro$ &(3/2,3/2)$^\ro$3 &&74.1082&74.1151 &73.8994\\
 57&                 $^1$D$_2^\ro$ &             &&74.1526 &74.1595 &73.9456\\
\vdots\\
 85&2s2p$^6$4d\ $^1$D$_2$ &(1/2,5/2)2&&84.0504 &84.0501 &83.9258\\
 86&   4f\ $^3$F$_2^\ro$ &(1/2,5/2)$^\ro$2 &&84.4770 &84.4789 &84.3462\\
 87&   4f\ $^3$F$_3^\ro$ &(1/2,5/2)$^\ro$3 &&84.4793 &84.4801 &84.3481\\
 88&   4f\ $^3$F$_4^\ro$ &(1/2,7/2)$^\ro$4 &&84.4853 &84.4839 &84.3522\\
 89&   4f\ $^1$F$_3^\ro$ &(1/2,7/2)$^\ro$3 &&84.4957 &84.4953 &84.3621\\[1.ex]
$\infty$&$\!$2s$^2$2p$^{5\,2}$P$_{3/2}^\ro\,\infty\,l$ &&92.760 &&---&&92.8398
\\\noalign{\smallskip}
\hline
\end{tabular}
\\[.7em]\ SS calculations with statistical model scaling factors
 $\lambda_{nl}$ = 1.3835 1.1506 1.0837\\
 \ 1.0564 1.0175 1.0390 1.0511 1.0177 1.0191 1.0755 in 1s 2s 2p\ldots 4f order.
\end{table*}

\end{document}